\documentclass[sigconf,screen,nonacm]{acmart}
\usepackage{xcolor}

\providecommand{\note}[1]{}

\AtBeginDocument{%
  }

\setcopyright{none}
\renewcommand\footnotetextcopyrightpermission[1]{}

\begin{document}

\title{SEAM: Situated Executable Associative Memory for a Co-located Community}

\author{Keru Wang}
\affiliation{%
  \institution{New York University}
  \city{New York}
  \state{NY}
  \country{USA}}
\email{keru.wang@nyu.edu}

\author{Michael Xu}
\affiliation{%
  \institution{Microsoft Research}
  \city{Redmond}
  \state{WA}
  \country{USA}}
\email{michaelxu@microsoft.com}

\author{Ken Perlin}
\affiliation{%
  \institution{New York University}
  \city{New York}
  \state{NY}
  \country{USA}}
\email{ken.perlin@gmail.com}

\author{Nebojsa Jojic}
\affiliation{%
  \institution{Microsoft Research}
  \city{Redmond}
  \state{WA}
  \country{USA}}
\email{jojic@microsoft.com}

\author{Andrzej Banburski-Fahey}
\affiliation{%
  \institution{Microsoft Research}
  \city{Redmond}
  \state{WA}
  \country{USA}}
\email{abanburski@microsoft.com}

\renewcommand{\shortauthors}{Wang et al.}

\begin{abstract}

In groups, knowledge is distributed among individuals, and how it is brought together depends on social ties. Taking inspiration from how a single mind  recalls and reacts to memories and current stimulus, we build a shared memory that does this for a community, prototyped as a social app, SEAM. Members post arbitrary moments; SEAM works out how fragments relate, distills knowledge about the group, and replies in a situated and associative way. Some of these memories act as instructions that can alter how the system behaves, making it programmable by its members. We deployed SEAM to 46 people in a co-located community for two weeks of open-ended use, and ran a follow-up workshop with 14. Members shared their days, interrogated the memory, and instructed it to track routines and run games; replies introduced them to people and subjects they had never encountered. SEAM points towards communities where knowledge finds people.

\end{abstract}


\maketitle
\section{Introduction}
When individuals live in a community, each comes to know something about its people, objects, activities, and places, encountering and understanding them in different ways~\cite{sutton2010distributed,barnier2008conceptual}. That knowledge lives in individual minds, and it becomes connected when the people holding it are connected and happen to share it. Rather than every member keeping a shared store of that connected knowledge, people tend to remember the connections better than the content, so recalling a memory often becomes recalling who holds it and going to talk to them, a division of memory across people that psychologists refer to as transactive memory~\cite{wegner1987transactive,liang1995group}. Because of this mechanism, a group usually remembers less than the sum of what its members individually know: knowledge travels only as far as the social graph that carries it, and it disappears with the people who held it~\cite{argote2000knowledge,orr1996talking}; detail is worn away each time something is passed from one person to the next~\cite{bartlett1932}; whatever does not serve the intention of the current conversation goes untold, even when it turns out to matter later~\cite{pillemer2003directive,conway2000construction}.

Besides other people, the shared artifacts a community keeps also help offload its memory and knowledge, and they strengthen the connections between members in different ways. For example, social media expands the social graph beyond the people a member already knows, linking them through common interests and keeping a copy of whatever they choose to share~\cite{leonardi2014socialmedia,levordashka2016ambientawareness}. Shared documentation works the other way around, accumulating knowledge under topics rather than under people, so a wiki, a search engine, or a question and answer forum lets a member look up what the community knows without first working out who knows it~\cite{kittur2008harnessing,morris2010what}. Recent research systems take up the same problem more directly. CHOIR turns the messages a research lab already exchanges into an organizational memory a chatbot can answer from, so knowledge is captured as a side effect of ordinary communication rather than through a separate act of documentation~\cite{lee2026choir}. Social-RAG retrieves from a group's own past interactions in order to ground what a model generates in that particular group with their shared memories~\cite{wang2025socialragretrieving}. Others place the agent inside the community as a participant of its own~\cite{seering2019beyonddyadic}, tag and summarize group chats so that what was said remains discoverable afterwards~\cite{zhang2018makingsense}, or surface the person who holds a piece of knowledge rather than the knowledge itself~\cite{mcdonald2000expertiserecommender,terveen2005socialmatching}. Each of these loosens the dependence on who happens to be connected to whom, and on what a member happens to say at the moment a connection is made~\cite{huang2026collective}. However, these systems either connect concepts in a predetermined way, as when a recommending system pushes material to people based on content similarities~\cite{eslami2015ialways,chaney2018algorithmicconfounding}; require a question before an answer can arrive, letting potentially relevant information go undiscovered whenever it falls outside the intent of the query~\cite{ackerman1996answer,dumais2003stuffi}; remain passive, waiting for someone to seek them out, and is forgotten along with the member who filed it~\cite{grudin1988why,rhodes2000justintimeinformation}; or represent the knowledge of the work rather than of the life lived around it, organizing information into topics, procedures and documents rather than around the particular people, places and objects its members actually move among~\cite{ackerman1996answer,lee2026choir,kittur2008harnessing,edge2024localglobal}.

Memory in an individual mind works differently: a recollection is reconstructed at the moment of recall out of bound fragments rather than replayed as a stored copy~\cite{bartlett1932,schacter1998constructive,konkel2009relational}; what surfaces depends on the cue and the situation as much as on the trace, so the same subject returns different moments depending on where and when it comes up, and a great deal of what we remember is never asked for at all~\cite{tulving1973encoding,godden1975context,berntsen2010unbidden,berntsen1996involuntary}. Memory exists in the service of action rather than of retrieval~\cite{glenberg1997what,pillemer2003directive}, so experiencing something in the past actually changes how people recall and react in the future. 

Hence, we ask what it would mean to give a community a memory that works in a similar way. What if we designed a shared mind for the community instead of a simple information holding artifact? Such a shared mind would wire its members' separate contributions into knowledge on its own, so recall would be distributed across many people rather than looked up from one of them. Could this system recall knowledge back from where somebody was and what they have just done, rather than from a query they had to think to ask? Could this shared mind be told what to do, so that what a member contributes changes how the memory behaves and not only what it holds? An individual interfacing with such a system would then come to know what their community knows, and have moments handed back to them as though they had lived alongside the people who experienced them.

Inspired by these thoughts, we elicited a set of design guidelines from how a single mind works with memory and we built SEAM (Situated, Executable, Associative Memory), a shared memory for a co-located community, prototyped as a social application. Members post arbitrary moments as text, photographs, video or speech, tagged with  time and location information. From these posts SEAM distills knowledge into a knowledge wiki: pages for the people, subjects, places, activities, events and relationships the community talks about, the connections that run between them, and the potential standing rules embedded in the users' posts instructing how the memory should react. When a new post arrives, SEAM reasons in an open-ended way about which shared memories might be related to this particular moment, combs through relevant earlier posts and wiki pages, and assembles the connections in situ to form a response. Since each recall is a reconstruction and reflection, it may produce new insights that SEAM  then records back into the  wiki, so the communal knowledge grows through use and not only through direct contributions. The knowledge wiki is accessible to members and allows for leaving comments, so that the shared memory can be inspected and refined in a way similar to that of individual knowledge. Crucially, in individual brains newly acquired information is stabilized via memory consolidation, which in SEAM we implement via  a consolidation agent that revisits the knowledge layer periodically, merging what belongs together, drawing out patterns spread across contributors, and acting on the comments left by members.

We deployed SEAM to a co-located community in an exploratory field study, to see what shared memories would form and what the members would do with them. Forty-six participants who live in the same city and work on the same campus used it over two weeks, contributing 345 posts, from which the system wrote 246 knowledge wiki pages and carried out 26 standing rules that shaped how it behaved. We report what emerged: what knowledge the system formed about the community and how it formed it, how it recalled and reacted, and what interactions arose among the participants. We then ran a follow-up workshop with 14 of them, to surface use cases that so short a deployment had no time to produce, to hear how participants had experienced it, and their suggestions on the system design.

This paper makes four contributions:
\begin{itemize}
  \item A set of design guidelines for a shared community memory, drawn from foundational work in cognitive science on how a single mind holds, recalls and acts on what it has.
  \item SEAM, a working prototype that instantiates those guidelines, including an interaction arrangement in which contribution and instruction share a single channel.
  \item An analysis of a two-week deployment and a follow-up workshop, reporting how knowledge was formed, how the system recalled and reacted, and which use cases emerged.
  \item A discussion of the design insights we draw from this and of the research directions it opens.
\end{itemize}

We hope SEAM is a step toward communities where knowledge is shared without being sought: where what one member works out reaches the next without the two of them ever having to meet; where a newcomer arrives already knowing the things everyone else knows; and where a community can act on everything it collectively knows without anyone having to gather it first.

\section{Related Work}
\subsection{How memory works in a mind and in a group}
\label{sec:related}

Memory in a single mind is organized rather than stored. What recurs across many experiences is extracted into knowledge about subjects, held apart from the episodes it was learnt from~\cite{tulving1972episodic,mcclelland1995complementary,ghosh2014schema}, and those subjects form a network whose links carry their own labels and strengths, so reaching one brings its neighbors within reach~\cite{collins1975spreading}. What may be bound together is not fixed in advance: the hippocampus relates elements that have nothing in common beyond having occurred together~\cite{konkel2009relational}. Retrieval over this network reconstructs rather than replays~\cite{bartlett1932,schacter1998constructive,conway2000construction}, what surfaces depends on the cue and the situation as much as on the trace~\cite{tulving1973encoding,godden1975context}, a great deal of it is never asked for at all~\cite{berntsen2010unbidden,berntsen1996involuntary}, and the whole arrangement exists in the service of action rather than of accurate recall~\cite{glenberg1997what,pillemer2003directive}.

A group holds the same material and has none of this. What it maintains instead is a directory of who knows what, so remembering often means remembering whom to ask~\cite{wegner1987transactive,liang1995group}, and the rest is spread across people, artifacts and the routines they work within~\cite{hutchins1995cockpit,sutton2010distributed,barnier2008conceptual}. Ordinary communication is what keeps the directory current, since watching colleagues talk teaches people who knows what and who knows whom~\cite{leonardi2014socialmedia}, mundane fragments accumulate into knowledge about others~\cite{levordashka2016ambientawareness,anderl2024ambientawareness}, and systems have long been built to leave that activity visible on purpose~\cite{erickson2000socialtranslucence}. As described above, the arrangement is bounded in three ways: it reaches only as far as the social graph and loses what departing members held~\cite{argote2000knowledge,orr1996talking}, it moves on present intention so an answer waits on a question being formed~\cite{ackerman1996answer,morris2010what}, and the artifacts that could carry the remainder stay passive, with the burden of preserving knowledge falling on the contributor rather than whoever benefits~\cite{grudin1988why,ackerman2000intellectualchallenge}.

It is the distance between these two arrangements that motivates our design. A community holds the fragments but not the network: nothing relates one member's contribution to another's, nothing distills what recurs across them, and nothing offers what is held at the moment it would matter. We therefore take the single mind rather than the shared document as the model for a shared memory.%

\subsection{Systems that help a group hold what its members know}

The tools a group already has divide along a line that matters here. Social platforms carry a great deal of what a community knows about itself, since people post meals, weekends and pets to colleagues and learn who is who in the process~\cite{dimicco2008motivationssocial,zhao2009peopletwitter}, and they surface things without anyone asking; but what decides which things is a prediction about engagement rather than any account of how the pieces relate, so what comes back is more of what you already attend to~\cite{eslami2015ialways,chaney2018algorithmicconfounding}, and the pile is a presentation of its authors rather than a record of a community~\cite{hogan2010presentationself}. Wikis, shared documents, search engines and question-and-answer forums invert this: they are organized by subject and a community can correct them, but they wait to be asked~\cite{kittur2008harnessing,ackerman1996answer,morris2010what}, and all of them inherit the asymmetry Grudin named, where whoever writes something down bears a cost that somebody else collects~\cite{grudin1988why}. One class arrives unasked but is ordered by engagement; the other is ordered by knowledge but only answers when asked.

Recent systems narrow this gap by surfacing information without a query~\cite{rhodes2000justintimeinformation,rekimoto1998augmentable,espinoza2001geonotes,zulfikar2024memoro}, linking fragments across captured moments~\cite{li2025omniquery}, and embedding memory systems within groups themselves~\cite{wang2025socialragretrieving,lee2026choir,zhang2018makingsense}. In that role, systems can become participants that shape group language and practice~\cite{seering2019beyonddyadic,hancock2020aimediatedcommunication,claggett2025relational}, while collective memory work highlights the tension between documenting the present and preserving it for unknown future uses~\cite{huang2026collective}.

Shared records are usually passive and fixed. End-user programming has long made behavior authorable through interfaces, trigger-action rules, and natural language ~\cite{maclean1990usertailorablesystems, ur2014practicaltriggeraction, li2019pumicemultimodal}. Communities can author executable policies too~\cite{zhang2020policykitbuilding, wang2024pikaempowering}. Recent work like Bonsai shows both promise and the difficulty of authoring in prose~\cite{elmalki2026bonsaiintentional, ko2011stateart,zamfirescupereira2023johnnycan}.

\subsection{How a machine memory is formed and maintained}

A system like ours also has to build the memory itself, deciding what to retain, which records belong together, and how to keep it usable as it grows. Agent-memory research touches upon much of this process:  Retrieval-augmented generation conditions a model on retrieved material~\cite{lewis2020retrieval}; generative agents keep a memory stream retrieved by recency, with reflection that writes higher-level inferences back~\cite{park2023generative}, to managed context~\cite{packer2023memgpt} and graph- or note-based structures that evolve with new material~\cite{gutierrez2024hipporag,edge2024localglobal,xu2025amemagentic}. This area is now surveyed systematically~\cite{zhang2025survey}. Two elements are load-bearing for us: reflection that writes conclusions back into a persistent store is the mechanism of our consolidation pass, and the two-store designs rest on the same complementary-systems argument~\cite{mcclelland1995complementary}.

We point this machinery elsewhere. In agent-memory systems the memory serves the agent's task performance, whereas in SEAM the memory is held for a community and recalled for its members, whose members can inspect, correct, and instruct it. In SEAM what has to be managed is whose experience reaches whom rather than how much context fits. Allowing contributions to become behavior also introduces authorship and prompt-injection risks~\cite{greshake2023notyou}, which we take up in Section~\ref{sec:discussion}. The nearest existing statement of this ambition is AI mediation of collective deliberation~\cite{tessler2024ai}, which mediates opinions where SEAM produces knowledge about a shared world.

\section{Design Principles}
\label{sec:principles}

A group holds the similar material as an individual mind but has a weaker mechanism for connecting it. This section takes the second half of that comparison seriously, asking what design insights we may learn from how a single mind works with its memory to create a shared memory system for a community.

\subsection{Defining central terms}
Before diving into details, we would like to clarify what we mean for the following terms in our paper:

\emph{Memory} is everything the system holds, the archive of contributions, interactions, together with what has been distilled from it. \emph{Knowledge} is only the distilled part, what the system has worked out about the people, places, things, activities, events and relationships the community has encountered. \emph{Recall} is the act of surfacing something from memory for present situation. \emph{Reacting} is acting on what has been surfaced, including generating a response as well as following the acting guidelines embedded in the user's interaction with the system. \emph{Remembering} as part of the system description is how the memory is stored in its form.

Two further terms are easy to conflate. \emph{Relational} is used in the sense of the relational-memory literature, the binding among fragments that recall reactivates~\cite{konkel2009relational}. \emph{Associative}, the A in the system's name, is used in a social sense instead: whose experience reaches whom. A memory is associative in our sense when a contribution from one person brings back what other people have encountered.

\subsection{How the knowledge layer holds what the group knows}

\textbf{\emph{D1. Knowledge is arranged by subject and linked open-endedly.}} As described in Section~\ref{sec:related}, a mind does not keep its knowledge inside the episodes it came from. What recurs across experiences is distilled into knowledge about subjects and held apart from those episodes~\cite{tulving1972episodic,mcclelland1995complementary,ghosh2014schema}, and the subjects form a network whose links are their relations, so reaching one subject brings its neighbors within reach~\cite{collins1975spreading}. The same two subjects can be connected for quite different reasons~\cite{konkel2009relational}, often several at once, and which of those reasons matters depends on the situation.

This suggests a distilled layer organized by subject, with connections preserved so recall can move between them. A similarity index cannot preserve relations that are not resemblances, while a graph of typed edges requires their meaning to be fixed in advance. We therefore leave relations in prose and link the subjects it mentions, allowing several reasons for a connection to coexist and letting recall interpret whichever matters in a given context.

SEAM implements this as a knowledge wiki of subject pages linked through resolved references, with claims grounded in the posts behind them (Section~\ref{sec:system}).

\textbf{\emph{D2. Knowledge sits above detail, and recall descends into it.}} Autobiographical knowledge is layered. General knowledge about periods and repeated events sits above event-specific knowledge, which is sensory and particular, and recalling an episode involves descending from the general frame into that detail~\cite{conway2000construction}.

A shared memory therefore needs both levels and a way to move between them. A distilled layer captures patterns that no single record contains, while the archive preserves the particulars lost in distillation. Each conclusion must therefore remain connected to the contributions behind it, allowing recall to descend from a general claim to the specific moments that support it.

SEAM keeps these layers separate but linked. Knowledge wiki pages cite the posts supporting their claims, leaving the underlying evidence directly accessible (Section~\ref{sec:system}).

\textbf{\emph{D3. Some knowledge forms only by revisiting the whole memory.}} Memory is not finished when an episode is encoded. Offline replay brings episodes into novel combinations, gradually extracting structure that no single episode contained~\cite{mcclelland1995complementary,lewis2011overlapping}. Consolidation therefore transforms memory rather than merely stabilizes it~\cite{dudai2015consolidation}.

A recall loop sees one situation at a time, and it can only draw the connections that situation calls for. Other knowledge appears only when the memory is examined as a whole: that two subjects are in fact the same subject, that a pattern spans contributors who never met, that a standing rule no longer applies to anything. None of these is reachable from a single stimulus, and none of them fits inside the time a member will wait for a reply. A shared memory therefore needs a slower process alongside the reactive recall, prompted by nothing in particular, whose work is to periodically reread and revise the whole.

SEAM implements this with a consolidation agent that revisits both layers on its own schedule, merging, generalizing, retiring, and recording its changes (Section~\ref{sec:system}).

\subsection{How recall should work}

\textbf{\emph{D4. Recall is reconstructed for the situation that asks.}} Recall assembles an episode rather than simply returns one~\cite{bartlett1932,schacter1998constructive}, and what it assembles depends on the cue and the context~\cite{tulving1973encoding,godden1975context}. Because this process recombines fragments, it can produce something that was never stored in that form~\cite{schacter2007constructive}.

Shared memory therefore should not be a direct lookup, but rather a reconstruction. The system has to decide what might bear on the present situation, retrieve relevant memories,  judge what has been established and what is still unresolved, and search again if needed. In doing so, recall can produce new knowledge and form new connections rather than merely surface old ones. This requires access not only to distilled knowledge but also to the underlying contributions, including along dimensions that were never their explicit subject.

Our system implements this with an iterative plan--retrieve--reflect loop over both the knowledge layer and the post archive (Section~\ref{sec:system}).

\textbf{\emph{D5. Recall fires without being asked.}}%

A present cue can surface related memories to mind before any question is formed~\cite{berntsen1996involuntary,tulving1973encoding}: a cup of coffee may evoke a memory of a great coffee shop, a friend who likes their coffee black, or an earlier morning mishap of having spilled a cup.

Group memory rarely behaves this way. A photograph of that cup may prompt no response at all, leaving related experiences scattered across the people who hold them. A shared memory should therefore treat the moment itself as the cue, bringing those experiences together without waiting for an explicit query.

Our system satisfies this by treating every contribution as a cue, whether or not it contains a question, and by recalling related moments contributed by others (Section~\ref{sec:system}).

\subsection{How the memory reacts}

\textbf{\emph{D6. Memory directs what happens next.}} Autobiographical memory is not only preservative, but also directive: a person's past experiences shape  future action~\cite{bluck2005tale,pillemer2003directive}. Its value lies in serving present behavior rather than merely maintaining an accurate record~\cite{glenberg1997what,suddendorf2007evolution}.

We borrow this directive function for shared memory. What members contribute should be able to change not only what the shared memory stores, but how it later recalls, responds, and organizes that knowledge. We therefore treat behavior as part of memory rather than as separate configuration, allowing instructions to arrive through the same language and channel as ordinary contributions.

Our system satisfies this by holding standing rules as ordinary pages written from ordinary posts, and by letting an execution change the reply or the knowledge wiki itself (Section~\ref{sec:system}).

\section{System}\label{sec:system}

Following the design principles above, we prototyped SEAM as a mobile web application for creating one shared memory for a co-located community. A user contributes a post as text, a photograph, a video or a spoken note, and every post carries its time and its GPS. The system connects that post to the earlier posts it relates to, and replies with what it has learnt about the community that may be useful or interested to the users' current moment. Users can follow up in the thread beneath any post, and the system reads those follow-ups back into the memory as well if there are valuable information . It keeps learning both from these interactions and from the reflections it produces while composing a reply, and what it learns is organized into knowledge wiki pages that users can read and leave comments on for the system to act on later. The application is a React Router 7 web app, and the shared memory is held in an Azure Cosmos DB database partitioned by document type, with media in Azure Blob Storage. The rest of this section describes how each part is designed and implemented.

\begin{figure*}[t]
  \centering
  \includegraphics[width=\linewidth]{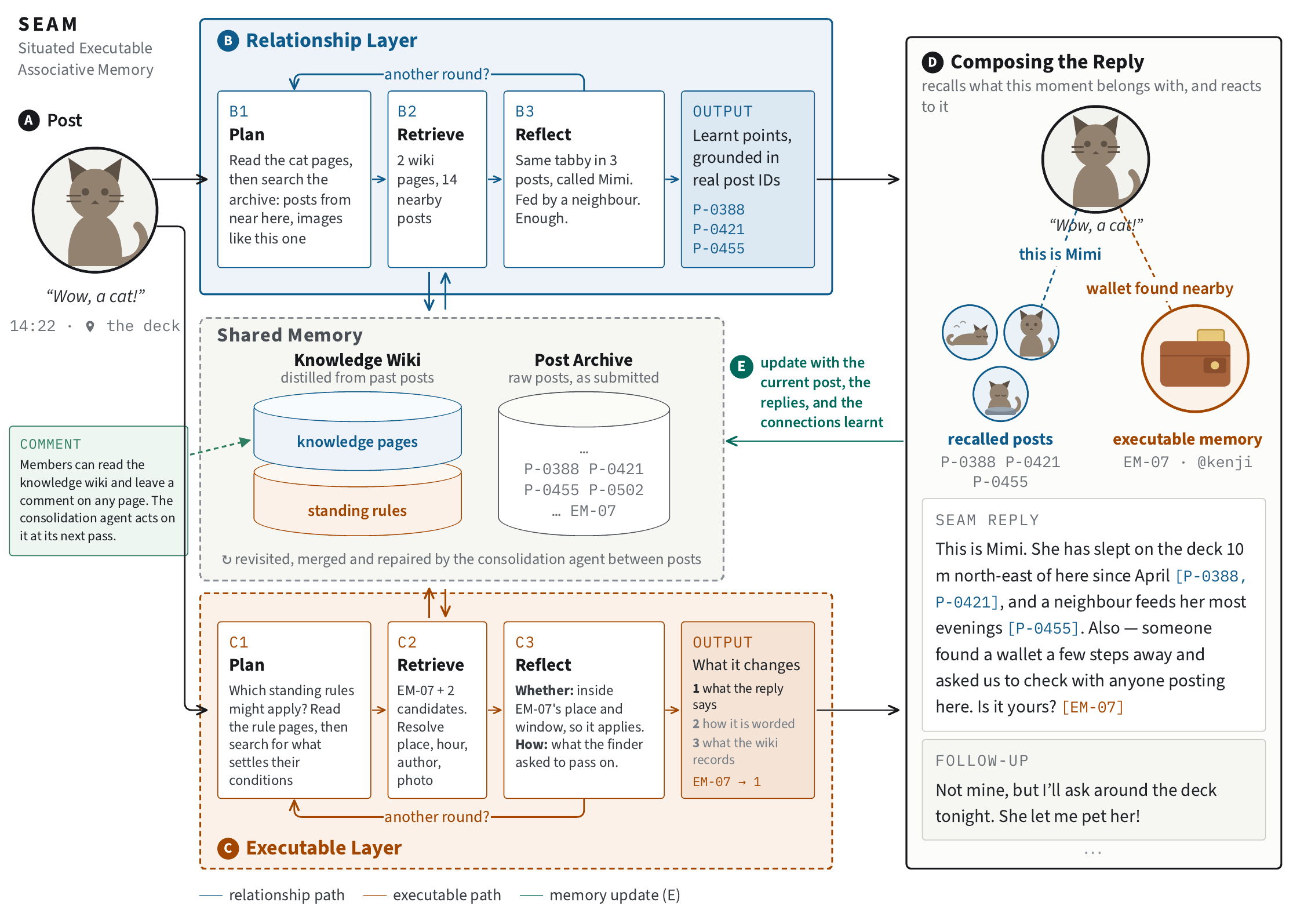}
  \caption{How SEAM handles one post, with an illustrative example rather than a case from the deployment. A post (A) enters both retrieval layers. The relationship layer (B) and the executable layer (C) run the same three steps: plan, retrieve and reflect, looping until each judges it has enough, and both read the same shared memory: a knowledge wiki distilled from past posts, and the archive of posts as submitted. They output reflections on the memory related to this moment, as well as context resolved regarding what actions to carry out and how to carry them out; The reply (D) is composed from both outputs with every claim grounded in the posts backing it, and the post, its follow-ups and the connections learnt are written back (E). Members can also read the knowledge wiki directly and leave a comment on any page, which the consolidation agent acts on at its next pass.}
  \label{fig:pipeline}
\end{figure*}

\subsection{The shared memory}

The memory holds two layers. The \emph{post archive} keeps each ingested post, where its speech is transcribed with Whisper, images and video frames are described by a vision model, each image is embedded with a CLIP vision-text model, and coordinates are resolved against places the memory already knows. The \emph{knowledge wiki} is what the system has distilled: pages for people, places, entities, activities, relationships, events and days, together with the standing rules elicited from the users' posts that the system uses to shape how it acts. A page contains prose with structured fields, and every cross-reference in it is a resolved knowledge page/user post identifier, so links are traversable in both directions while the reason for a link stays in the text rather than in an edge label (D1). A page's claims always refer to the posts that suggest them, so the claims are more trustworthy and ground truth is always one link away (D2).

Each post also receives a page linking it into this graph, and those pages are kept out of the index the retrieval stage is shown, since posts are directly searchable. A standing rule is a page linked like any other, but it is specifically recording how the system should act suggested by the users' posts. Its knowledge page records its action trigger condition, the action, its scope, whether it is in force, and the action it has taken so far.

\subsection{Two retrieval layers over one memory}

There are two retrieval layers: one is to figure out what are the relevant memory and knowledge about the current post; the other is to figure out what action the system may need to take in react to this post. Both layers work the same way, in three steps that can repeat (D4). In \emph{plan}, the model is given the post, the wiki's index of pages and a set of search tools, and produces a plan naming which pages to read and which searches to run over the archive: what to match on, whether to constrain by time, author or place, and how much weight visual similarity should carry for this particular moment. In \emph{retrieve and read}, the named pages and the search results are read together, and what has been learned is stated as a set of points, each tied to the post that supports it. In \emph{reflect}, the model judges whether that is enough, naming what has been established, what is still missing, and which links or further searches would settle it; if more is needed it plans again, now holding what it has just learned. Both layers ran a smaller, faster model (GPT-5-mini), since their work is search and judgment across many passes rather than composition.

For example, if we ask the system where to find Mimi, the first plan may be generic, because the name means nothing yet, but reading the relevant pages mentioning Mimi establishes that the user is referring a black cat, is usually seen near one part of the campus, and belongs to a particular household. The second pass can then search the archive for descriptions of such a black cat, constrain the search to that area, and run visual similarity against a photograph it has just found, so a name in a question becomes a set of sightings by people who never met. This is also why recall can produce knowledge the memory did not already hold: reflection states conclusions that were written nowhere before it ran, that this animal is the one in an earlier photograph, or that two accounts describe a single occasion. Connections are reconstructed at recall rather than looked up (D1, D4).

We designed six tools that can search over the archive: filters on time, on author and on location, exact keyword match, keyword similarity over an Azure OpenAI text embedding deployment, and visual similarity over the CLIP image embeddings. From any page the model can also follow resolved links to related pages. This is the descent D2 requires, and it works along dimensions that were never a post's subject, so a moment can be reached by when it happened or by what it looked like rather than only by what it was about (D2, D5).

The \emph{executable layer} runs the same three steps and spends them on resolving if users want the system to act in a certain way when reacting to this post (D6). In the knowledge layer, memories that carry potential action guidelines are write as standing rules in the knowledge wiki, and the executable layer's job is to compare the current post against the standing rules to resolve if an action may be triggered and the context needed to decide how to act in this particular case. We design the system in a way that the memory shared by users can not only shape how the system response to the user, but also how it update its knowledge layer to make the system more open-ended.

\subsection{System response}

The system composes its reply from the outputs of the two retrieval layers: the memories judged relevant to the current post and any standing rules that should apply. It groups the supporting posts by the relation it inferred, carries out any applicable instructions, and states its conclusions in prose with inline citations. This composition step uses the larger model (GPT-5.4), where the deeper reasoning is most useful.

Citations always resolve to posts and never to wiki pages. Firstly, SEAM exists to connect what the community has already shared rather than to produce claims of its own, so everything claim has to be grounded in something a member actually contributed to shared memory. Second, posts are immutable once submitted, whereas wiki pages are continually rewritten, merged, and renamed as the memory evolves; citing the distilled layer would therefore make references unstable.

\subsection{Consolidation agent}

After replying, the post, its follow-ups and the connections just established are written into the wiki by the same larger model, so reasoning performed during one recall is available to the next rather than recomputed (D4).

Separately, a consolidation agent revisits the whole memory on a schedule, in incremental and full-sweep modes, with read and write tools over both layers and no latency budget to meet (D3). It does what a stage seeing one post at a time structurally cannot: merging pages that describe the same subject, finding patterns spread across contributors, retiring rules whose conditions have passed, acting on the comments members have left, and repairing what the per-post stage got wrong. Each pass writes prose notes recording what it changed, what it deliberately left alone and what it could not settle, which is what makes its behaviour analysable at all.

\subsection{What users see, and how they speak to it}

Users address the memory through three channels, and all three write into it. A \emph{post} is a new moment, and it is the primary channel. A \emph{follow-up} is a message in the thread beneath the system's reply, read back into the memory rather than treated as chat, which is where questions, corrections and further instructions arrive (D6). A \emph{comment} is a note left on a knowledge wiki page, which the consolidation agent reads and acts on at its next pass (D3). Correction therefore has no dedicated mechanism, because disagreeing with the memory uses the same channels as adding to it.

A reply has two parts. Above sits the original post, and dashed spokes fan out to the memories it was joined to, each spoke labelled with the relation the system drew. Where a spoke carries several posts they share one circle and cycle through it, with a count showing how many (see Figure~\ref{fig:screens}; holding the circle expands the group, and selecting a member opens that post directly. Below, the reply is written as prose, with every claim carrying inline citations to the posts it rests on, so a reader can move from a statement to the moment behind it in a single step. The graph shows which memories were reached, and the prose says what the system made of them.

The knowledge wiki is browsable as a set of web pages with a hierarchical menu, and because every reference in a page is a resolved link, users navigate what the system has concluded about their community the same way the retrieval stage does. We make this layer visible because its contents are about the people reading it: a member should be able to see what the memory has worked out about them and their community, and say so on the page when it is wrong. 

\section{User Study Setup}

We ran two studies with different purposes. A field deployment asked what a shared memory does when a real community uses it with no task assigned, and what people use it for. A follow-up workshop asked what such a memory should become, covering uses the deployment never reached, and collected feedback on the design and its limits. Both were approved through the host organisation's research ethics review. An AI content filter and a post-reporting feature were built into the system for the deployment: a reported post is withheld immediately and sent to the researchers for review.

\subsection{Field deployment}

We recruited by email from a co-located community whose members work on one campus and live in the same city, and 46 people signed up (31 male, 15 female, age ranging from 24 to 55 years old). Each was contacted by a researcher who explained the system and its features, including that posts can carry instructions the memory will act on, where to read the knowledge wiki and how to comment on it, and that replies could be rated in the app. Participants registered with their organisational email, signed a consent form and agreed to terms of use covering risky and sensitive content. They were given no task and used the app freely for two weeks.

System replies could also be rated in the app. A rating is a thumbs up or down with an optional reason. A thumbs up offers \emph{relevant memories}, \emph{correct answer}, \emph{useful discovery}, \emph{clear connections} and \emph{easy to use}; a thumbs down offers \emph{incorrect answer}, \emph{irrelevant memories}, \emph{missing memory}, \emph{confusing labels}, \emph{hard to use}, \emph{too slow} and \emph{privacy concern}. Either can be given as \emph{other} with a reason written in.

\subsection{Follow-up workshop}

After the deployment we invited interested participants by email and fourteen took part (W1 to W14), all of whom had used the system, so their proposals come from people who had lived with it rather than from an outside panel, and they are self-selected for interest rather than sampled. The session ran on three prompts: what they would use a memory of this kind for given more time, what they thought of the design, and what they would change about the version they had seen. They were asked to think beyond what the prototype could do, wrote their responses individually for thirty minutes, then discussed them together for thirty more, so the record holds both considered written proposals and the discussion that followed. The workshop was deliberately generative rather than evaluative; its purpose was to reach the uses and objections that one community could not surface on its own.

\subsection{Data preparation and analysis}

Our analysis draws on the posts as submitted with their speech transcriptions and visual descriptions, the replies, the in-app ratings, the follow-up messages and page comments, the knowledge wiki, the consolidation agent's own reflection notes, and the workshop's written responses and recording. Counts of what the system produced come directly from that record. The classifications we report are hand coded: every one of the 345 posts was labelled for the four modes of use, and the questions, instructions, follow-ups and workshop proposals were coded thematically with categories developed from the material rather than fixed in advance. To respect participants' privacy, real names, businesses, landmarks and buildings are generalised throughout, interface figures are redrawn rather than reproduced as screenshots, and the example replies in Figure~\ref{fig:screens} are reconstructions rather than participant data.

\section{Findings}\label{sec:findings}
\begin{figure*}[t]
  \centering
  \includegraphics[width=\linewidth]{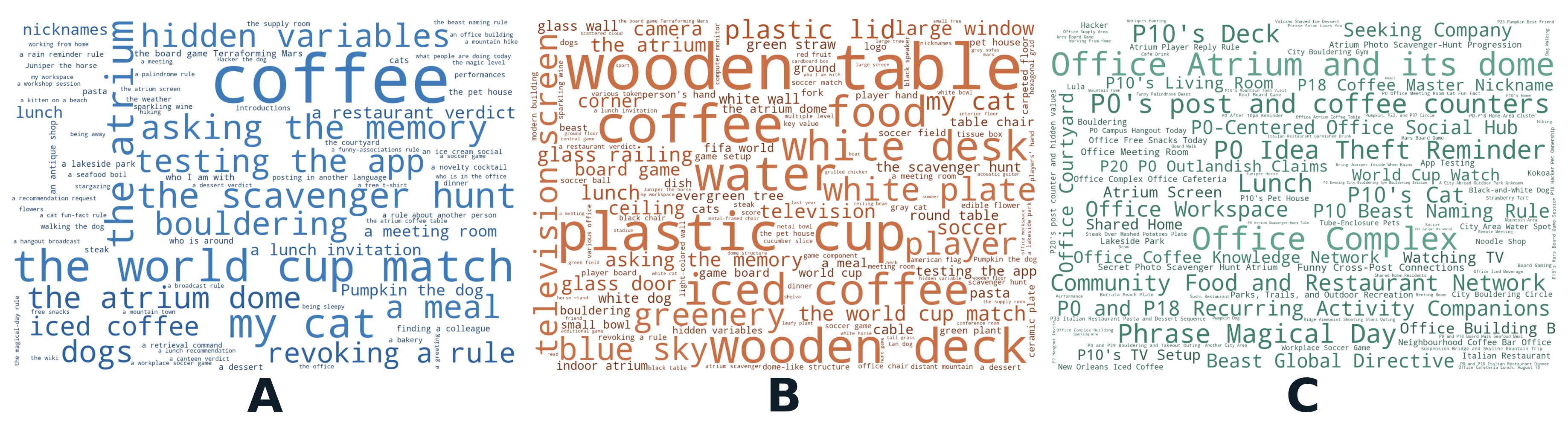}
  \caption{Word clouds of the subjects in the 345 posts: (A) what the poster was focused on sharing, (B) what the system captured from the same post, and (C) what the knowledge wiki recorded. Text size follows the number of posts behind each subject, on a square-root scale within each panel.}
  \label{fig:clouds}
\end{figure*}

Over two weeks, 46 participants who live in the same city and work on the same campus contributed 345 posts to SEAM. We examine four things in turn: what emerging use cases the participants created, how it recalled and reacted when someone posted, the knowledge SEAM distilled from those posts, and what the participants would want from a memory of this kind next. Figure~\ref{fig:clouds} reveals more detailed content of the participants' posts.

\subsection{What People Did With It}

\begin{figure*}[t]
  \centering
  \includegraphics[width=\linewidth]{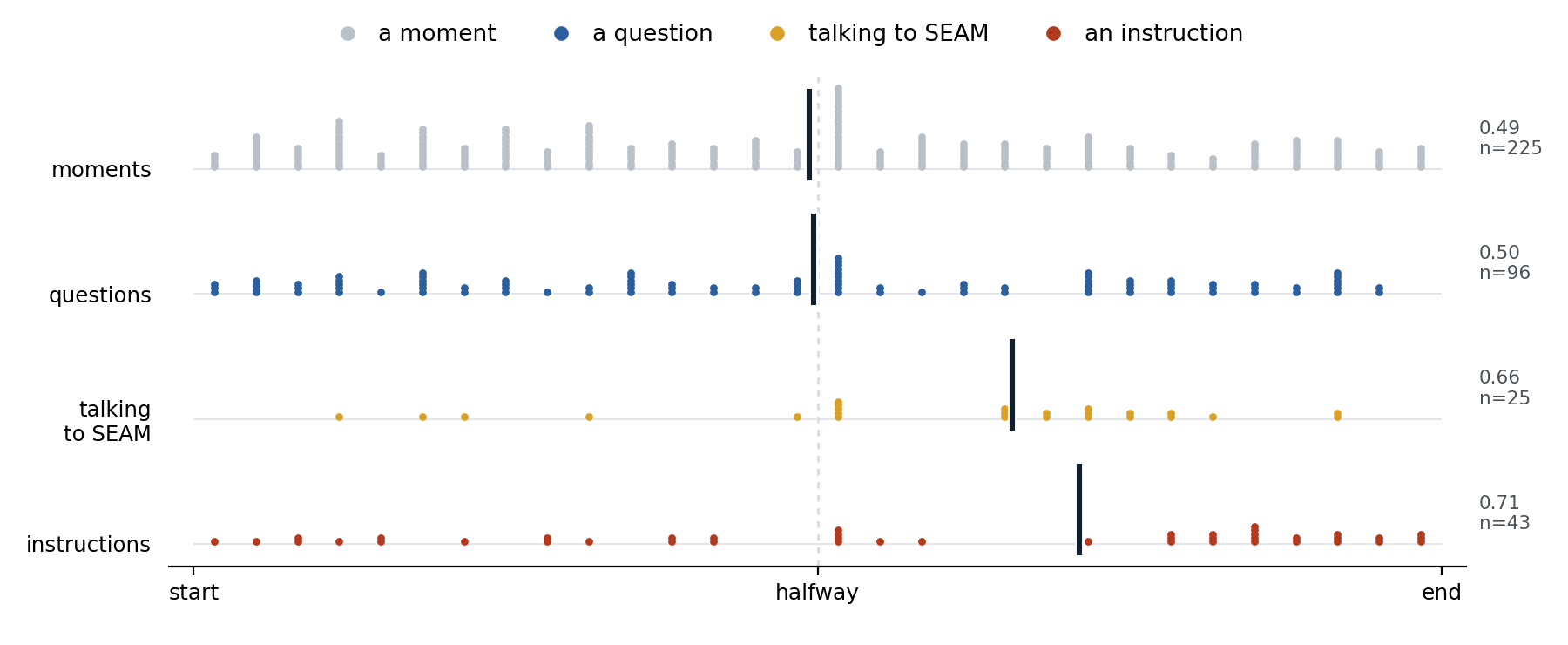}
  \caption{The four kinds of post, distributed across each participant's own posting period.}
  \label{fig:modes}
\end{figure*}

Participants shared 345 posts. We code each by what it does into the following four non-overlapping  kinds: 

\paragraph{\textbf{Moments.}} 225 of the 345 posts, from 37 of the 46 participants, simply record something from the day with no question attached and nothing asked of the system: for example, their pet on a roof, the food they had, a hiking session, a view from the park. This is the most common use and the most evenly spread, sitting across the whole of a participant's run in Figure~\ref{fig:modes}.

\paragraph{\textbf{Questions.}} 96 posts, from 28 of the 46 participants, put a question to the community. These include direct requests for information about the buildings, where to find a facility, open social invitations such as who is around for dinner, who left a T-shirt on their desk, who a newcomer should talk to if they want to meet the outdoor enthusiasts, etc. They are spread almost as evenly as moments across a participant's run (Figure~\ref{fig:modes}), and are the second most common post type.

\paragraph{\textbf{Talking to SEAM directly.}} 25 posts, from 15 participants, seems to be talking to the app directly through their posts instead of to the people in the community. Some test what it can do, asking it to guess who the poster is with; others treat it as an agent to be consulted about itself, asking where to find the knowledge wiki, whether it can look something up for them, or to explain how a feature works. Workshop participants described this shift directly: seeing the system compose replies and carry out executable memories moved it, over time, from a social app toward an agent they were talking to (W8, W10). The posts sit accordingly in Figure~\ref{fig:modes}, at a median of 0.66 through a participant's run, later than moments and questions but well before instructions, as they get more familiar with the system and start to testing it more.

\paragraph{\textbf{Instructions.}} 43 posts, from 11 participants, carry an action that shapes how SEAM behaves. Some broadcast: for example, a party invitation to anyone posting within a set window, a reminder not to stay late for anyone posting after ten at night, a note to whoever is nearby that somebody has brought snacks to share, a warning that a wallet had been lost in a park and that people should watch their own; Some govern how the system speaks, sharing that ``funny replies would be welcome in the community'', or fixing a naming convention for the community such as telling the community one should be referred to ``the coffee master'', ``never call my cat a cat but always the beast'', etc; Some reach into the knowledge wiki, which is where posting becomes more like direct ``programming the system'' (W2), as when a participant asked it to keep a counter of their coffee posts and total what they had spent over the weeks. Some are playful, asking SEAM to write outlandish claims about another participant onto that person's page, which it did while recording that the claim came from a member rather than from evidence. And some are written against each other, telling the system to stop an existing rule, or stacking a further condition on top of one already in force.

The most elaborate case is a scavenger hunt one participant set up in four posts over ninety minutes, with a set of chained instructions on how they would like the system to act. They asked SEAM to invent the game itself, choosing 3 target objects in the atrium and writing a hint for each: and the system could do that because it already held a page for that place listing what is in it, so a dome, a large screen and a coffee table became the ordered targets without anyone naming them. A second instruction recruited players, marking anyone who posted within twenty metres of the atrium that day as a participant and letting SEAM to tell them so in their posts' reply. A third post put the state in the knowledge wiki, letting the activity page record the players and their scores, while each player's current step and next hint are written onto that player's own page when they captured the relevant objects in the photo they posted. A fourth reads the finish order back, congratulating a returning player and naming their place in it. Progression is a rule of its own, firing on any post from the on-boarded players whose photograph is captioned with one of the three objects and then working out whose move it is. The rules expire as the game does: the recruitment rule was revoked an hour after it was written, and the player-reply rule lapsed at the end of that day. Five other participants were recruited into it and played it through the replies alone, having been told the rules by nobody.

Because this is an unique feature our system carries, it is the least common of the four kinds and the latest to appear, at a median of 0.71 through its author's own run (Figure~\ref{fig:modes}).

\subsection{How the Memory Recalls and Reacts}

This section examines what happens when a participant contributes a moment: how the memory answers, what kinds of connection it reconstructs, and how participants respond to what comes back.

\subsubsection{What a reply is made of}

\begin{figure*}[t]
  \centering
  \includegraphics[width=\linewidth]{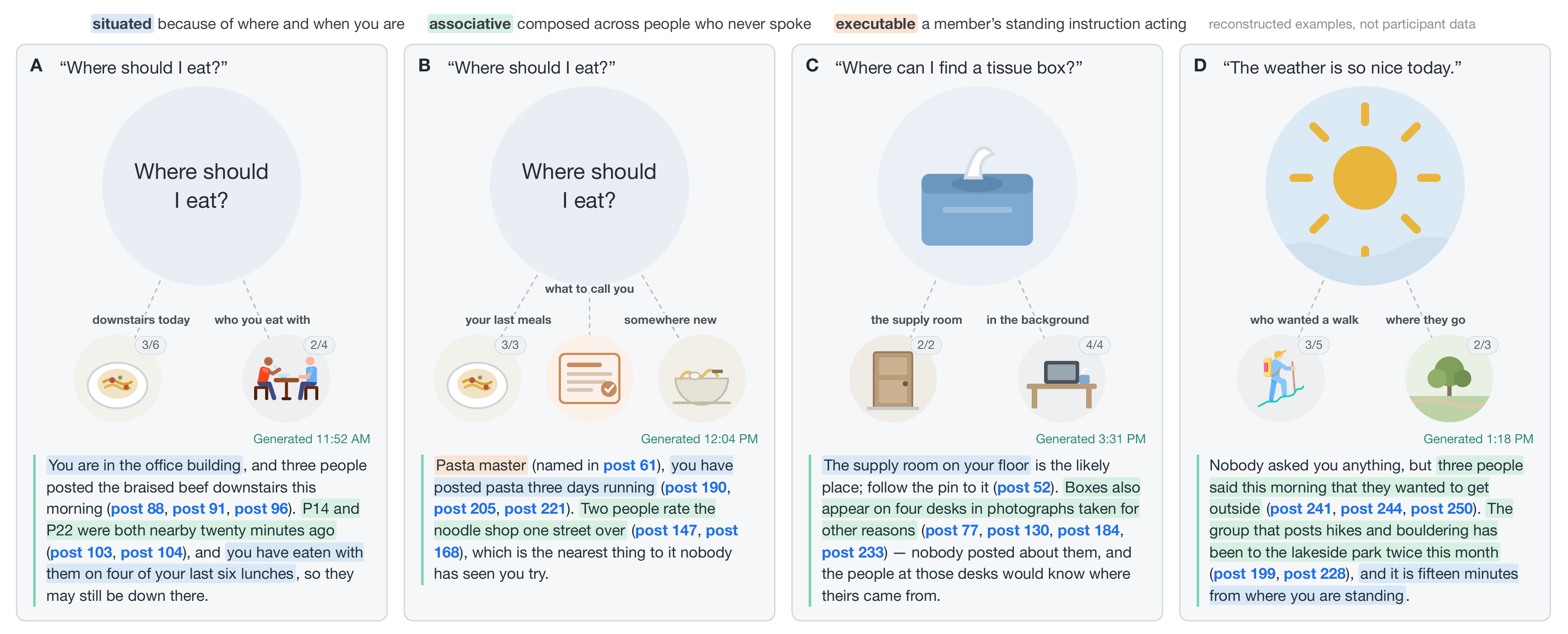}
  \caption{Four replies, reconstructed rather than taken from the deployment: (A) and (B) the same question answered differently for two members, (C) an object found in the background of photographs taken for other reasons, and (D) a reply to a post that asked nothing. Highlighting marks what makes each part situated, associative or executable.}
  \label{fig:screens}
\end{figure*}

A reply is assembled from both layers and shaped to the moment in front of it, and most carry three qualities at once rather than one at a time. It is \emph{situated}, taking present circumstances into account in deciding what is worth surfacing; \emph{associative}, connecting the contributor to the rest of the community rather than only answering them; and \emph{executable}, carrying out instructions the community has left in the memory when the moment satisfies their conditions. Figure~\ref{fig:screens} shows four replies, reconstructed rather than taken from the deployment, with the three qualities marked in the text of each.

\paragraph{\textbf{Situated.}} The same question returns different answers to different people (D4). Panels A and B of Figure~\ref{fig:screens} are both ``Where should I eat?''. A is answered from where the asker is standing and who they usually eat with: three people posted the dish downstairs that morning, and two of the people this member eats with were nearby twenty minutes ago, so they may still be down there. B is answered from the asker's own history instead, three days of pasta, and points them at a noodle shop two other members rate as the nearest thing to it they have not tried. Circumstance is not a filter applied after retrieval but part of what makes something worth recalling at all.

Because nothing is discarded at capture, relevance is not limited to what anyone meant to record (D2). Asked where to find a tissue box (Figure~\ref{fig:screens}C), the memory named the supply room and then added four desks where boxes sit in the background of photographs taken for entirely different reasons, suggesting that the people at those desks would know where theirs came from. That background is most of what the system holds: participants were pointing at 206 subjects while their posts carried 2,075 (Figure~\ref{fig:clouds}A and B). Nobody could have known at the time which of them would matter, which is what lets the memory serve as documentation of a workspace nobody set out to document.

\paragraph{\textbf{Associative.}} An associative reply brings separate people's experience into contact (D1). Mostly this is not a referral: someone photographing an animal inherits the name another participant gave it, and someone looking for an object is told where it was last seen. What moves between people is the knowledge itself rather than a pointer to whom to ask. It does not wait to be asked, either (D5). Panel D of Figure~\ref{fig:screens} is a post that asked nothing at all, ``the weather is so nice today'', and the reply returns three people who had said that morning that they wanted to get outside, together with the park the community's outdoor group has been to twice this month and how far it is from where the poster is standing. Participants described this less as searching than as inheriting an experience: looking at something and immediately knowing what it is called, as though you had been there before when you never had.

\paragraph{\textbf{Executable.}} Unlike the first two qualities this one has an expectation attached: when a rule fires there is a right answer about what should have happened, so it can be checked rather than only described (D6). Instructions usually act inside a reply that is doing something else anyway, as in Figure~\ref{fig:screens}B, where the member is addressed as \emph{pasta master} because a third participant had written a rule naming them that.

Simple instructions are carried out reliably. All seven rules governing how the memory refers to a person or animal fired whenever their conditions were met, as did rules keyed to a time window or a location, since satisfying those requires only comparing the post against a stated condition.

Complexity arrives with ambition. Some instructions are triggered not by the post but by the reply the memory is about to give, as with a phrasing rule that applies to any response mentioning a particular subject. Others need several steps: running the scavenger hunt meant establishing whether a poster was playing, resolving how far through they were, choosing the hint for that stage and updating their standing. Where a rule requires resolving state before acting, steps are sometimes dropped, and participants who understood the conventions occasionally repaired this in the reply thread. A different failure is misreading who should act: asked to tell everyone who posted in a room a cat joke, the system first told later posters that they should tell the joke, and corrected course only when the participant said so.

Most replies carry all three qualities together, which is how they were designed to work: a recall shaped by circumstance, reaching for a connection nobody requested, and carrying out whatever the community has asked of the memory when this kind of moment arrives.

\subsubsection{\textbf{How participants responded}}
\paragraph{\textbf{Suggest correction on connections through the reply feature.}} 71 posts drew at least one follow-up in the thread, 155 in all from 21 participants, unsolicited and several times more numerous than the ratings. Around a quarter correct the memory or add to the original post: an animal it recorded as a cat was a pet rat, a place it named as one park was another, two accounts belong to the same person, a keyboard it retrieved was a computer keyboard where the poster meant a musical one. Corrections, challenges to whether a recall was complete, and questions about how the memory knew what it claimed together account for about two in five, so participants used the thread less to query the memory than to argue with it and repair it. 

\paragraph{\textbf{Ratings.}}
Rating a system response was available but never enforced, so the ratings that exist mark the occasions on which someone chose to react. Twelve participants left 27 ratings, of which 23 are positive: 17 of 19 on generated replies and 6 of 8 on answers in a thread. What the positive ones cite is the more informative part: relevant memories in eight, clear connections in six and useful discovery in four, against correct answer in two. The reasons participants reached for describe discovery and connection rather than correctness, which is consistent with a memory that surfaces rather than a service that answers. We treat these as an indicator of what prompted someone to react rather than as a measure of quality.

\subsection{What the Memory Came to Know} %
The system distills knowledge about the community and writes it into a knowledge wiki from user's interaction. We first look at what content the wiki came to hold, then analysis how it is formed, and finally at what the consolidation process contributes to it.

\subsubsection{\textbf{What the knowledge layer covers}}
By the end of the deployment the memory held 246 knowledge pages: 65 entities, 59 places, 43 people, 26 standing rules, 19 activities, 19 events and 15 relationships. Pages nest, so the layer reads as a menu rather than a list. Places sit under five complexes and seven regions, entities are grouped further into animals, accessories, appliances and more, and every activity carries a domain the system chose for itself, such as sport, work, outdoors, food and games.

What a page holds is not a set of keywords gathered from the posts but an account of what goes on. An office building assembles the equipment visible in the background of photographs, who is usually there and what recurs, amounting to live documentation nobody set out to write, and an open social invitation is marked closed on its page when the poster shares moments from the meeting shortly afterwards. The higher-level pages are interpreted from how separate fragments fit together. Several accounts posting daily life from the same coordinates in domestic surroundings become a shared residence although nobody said so; one participant is recorded as the community's social hub on the strength of cross-mentions and shared occasions; and people whose posting histories keep returning to bouldering and hiking are gathered into a sub-community of likely outdoor enthusiasts. This can also be shown in Figure~\ref{fig:clouds}, as in (C) it holds much higher-level topics than contents from the posts directly.

Because members can instruct the memory, the layer also holds subjects that exist inside SEAM rather than in the physical community, such as the counters participants asked it to keep of their own posts and coffees, and the scavenger hunt the community invented with SEAM described earlier. More detailed description of each type of wikis is available in supplementary material.

\begin{figure}[t]
  \centering
  \includegraphics[width=\linewidth]{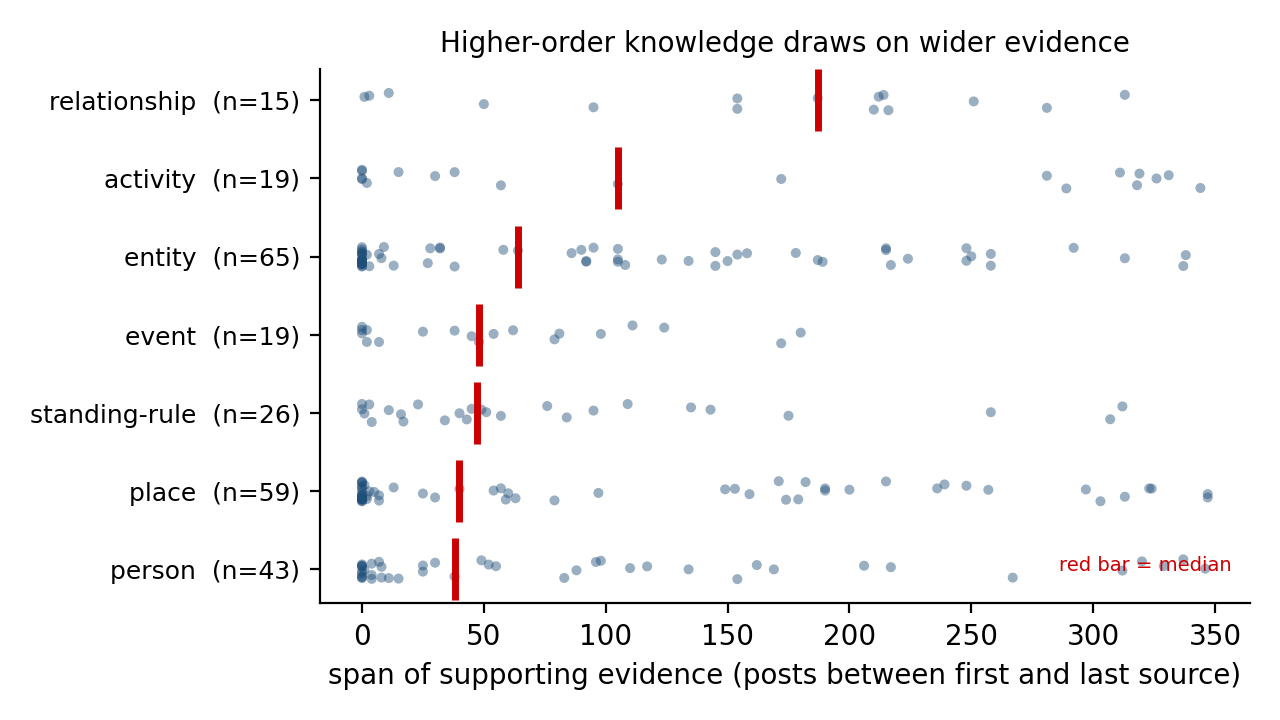}
  \caption{How far apart the posts behind a knowledge page lie, by page type, measured in posts rather than in days.}
  \label{fig:span}
\end{figure}

 From Figure~\ref{fig:span}, it also shows that page types differ in how far apart their supporting posts lie over time: relationships span furthest, a median of 187 posts, because a tie between people is read off behaviour repeated across a long stretch rather than stated once, and recurring activities follow at 105, where events sit at 48 because an occasion has a short window and standing rules at 48 because most were written for a trigger window that then passed. Entities, places and people are spread far more unevenly, anywhere from zero to nearly the whole deployment, since a subject's span tracks how popular it became and how long it stayed active.

\subsubsection{\textbf{The knowledge wiki's growth pattern}}

\begin{figure*}[t]
  \centering
  \includegraphics[width=\linewidth]{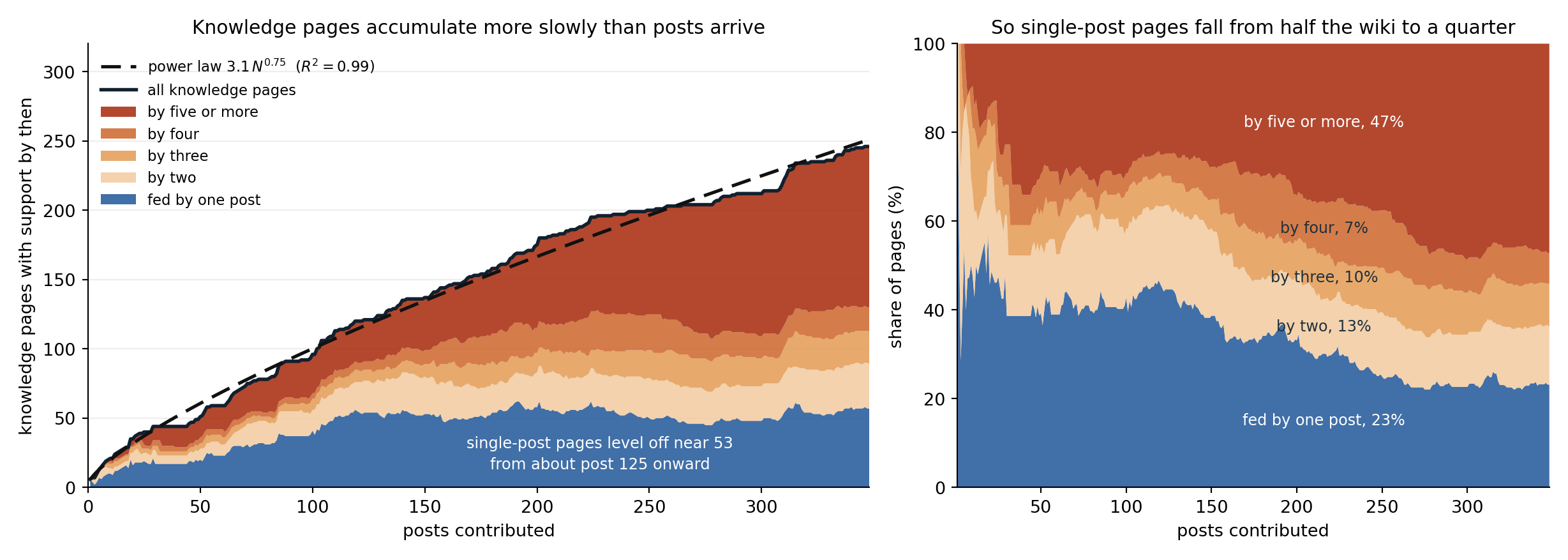}
  \caption{How the knowledge wiki filled in as posts arrived, with pages coloured by how many posts stand behind them: (A) as counts, against the fitted power law, and (B) as shares.}
  \label{fig:growth}
\end{figure*}

Figure~\ref{fig:growth} shows how the knowledge layer filling in as posts arrived. Each page is dated to the earliest post that supports it, and the line counts pages with at least one supporting post at that stage against posts contributed. The bands split that total by how much stands behind a page at that moment, pages fed by one post, by two, by three, by four or five more, and the right panel shows the same bands as shares. 

Growth is sublinear: the number of distinct subjects rises as a fractional power of the number of posts, $K \approx 3.1\,N^{0.75}$. This is Heaps' law~\cite{heaps1978}, first observed as the relation between the length of a text and the number of distinct words in it, and an exponent below one means each new post is less likely than the one before it to raise a subject nobody has raised yet. A group drawing on an unbounded supply of unrelated subjects would grow close to linearly; one that keeps returning to a shared set grows sublinearly, and the exponent says how strongly. Across the corpora measured in the same way by~\citet{tria2014dynamics}, the exponent is 0.45 for Gutenberg texts, 0.56 for the artists people listen to and 0.68 for song lyrics, against 0.77 for Wikipedia editing and 0.78 for social tagging. Ours, at 0.75, falls with the second pair and well away from the first. We read that as a sanity check: the low-exponent corpora are collections whose entries do not build on one another, and a memory whose members were each keeping a private diary would resemble those. That ours resembles the collective cases is some evidence that people were adding to subjects other people had started rather than accumulating in parallel.

The bands say how much stands behind a page at each point, and two things move. The five-or-more band widens steadily, so a post made late in the deployment arrives at a subject that already has a history and the reply it receives has more to draw on; since the system surfaces related material only where it finds some, this is the memory growing rather than the retrieval loosening. At the other end, the share of pages resting on a single post falls from about half to a quarter while their count holds near 53 from about post 125 onward, so single-post pages are a turnover rather than a dead tail: new subjects keep opening at roughly the rate older ones acquire a second contributor. Some of those second contributors arrive because people feed off one another, seeing a reply tie their own post to somebody else's and then posting something adjacent to it, which is why a large share of pages stops being single-post so early.

Taken together, the curve and the bands say that distilling posts into subject pages produced a memory that widened like a collective wiki rather than a pile of private diaries, and that deepened as it went, so each later post arrived somewhere that already had something to offer it.

\subsubsection{\textbf{What was done in the consolidation}}

A consolidation agent ran over the whole corpus on a schedule, reading the knowledge layer as a body rather than one post at a time, and wrote notes at every run on what it found, what it changed, what it deliberately left alone and what it could not settle. Across 205 runs it recorded 184 pieces of emergent knowledge, 251 repairs, 27 merges, 297 deliberate non-actions, 185 statements of uncertainty and 65 responses to comments participants had left on pages.

\paragraph{\textbf{Most of its work was repair and reorganisation, and corrections were weighed rather than applied.}} It resolved missing links, removed duplicate subjects and fixed what the per-post stage had got wrong. When one participant asked for a hierarchy over the animals, the pass generalised the request and gave every entity a domain and type, every activity a domain, every relationship a kind and every place a region, noting in its own record that it had applied the request corpus-wide rather than only to the page it was raised on. Corrections were not taken at face value: for example, a blurry photograph had been classified as a cat, and another participant commented that it was a pet rat. The pass accepted that the original label was overconfident but refused to assert the correction, recording two reasons: the photograph could not settle the species, and the commenter was not the person who had posted it. The page became an unidentified small pet. The agent is being careful to make sure the knowledge base is safe and trustworthy.

\paragraph{\textbf{It found the relationships no single post could state.}} Nine of the fifteen relationship pages were written during sweeps rather than in reply to any post: two participants' recurring companionship assembled from outings months apart, a shared residence inferred from several accounts posting domestic life from the same coordinates, and the interest networks around coffee, food and the outdoors, each gathering people whom no single post had connected. None of them was drawn during a recall. Each needed the memory to be read as a whole, after the fact, which is what D3 asks for and what a stage seeing one post at a time structurally cannot do. None required a participant to ask either, and once written the pages became available to recall, so a relationship the memory had worked out on its own could later be surfaced to somebody looking for who to talk to.

\subsection{What People Would Do With It}

We summarized what people would like to do more with the system if given more time, to show potential future use cases SEAM may apply to.

\subsubsection{Assembling a situation nobody can see whole}

W9 proposed a rescue or escape-room task in which each person explores a different area, every observation feeds one memory, and the team stays synchronised without anyone stopping to report. W7 described neighbourhoods where people share which route has become risky or where break-ins are happening, information that turns over daily and currently scatters across social-media groups, and wanted entering a flagged area to be what surfaces the warning. This is the class of settings where many people each hold a fragment of one situation and nobody holds the whole. The prototype already supplies the awkward part, since posting costs nothing and place and time are captured without being stated, and recall assembles across contributors who never coordinated. What it lacks is delivery: both proposals need a recall to reach somebody who has not posted, and both fail by being passive rather than by being wrong.

\subsubsection{A connecting layer under what people already produce}

Nobody asked for another place to post. They asked for a memory attached to where they already are (W9), and named where accumulated knowledge has become unreachable: ``I have found Reddit and X especially difficult to search. Everything on X is pretty much lost'' (W4), with the discussion that once sat on public question sites having moved into chat servers that cannot be searched at all. W2, having seen the wiki, described the system as an agent building its own forum out of posts nobody had categorised, and wanted those categories to be the communities. The architecture does not depend on the social application: anything with a stream of posts and a situation to be answered in can sit under it, a chat server whose history is unusable, a team whose decisions are spread across threads, an institution's records. The requirement is a stream, not a new destination. That is also the limit of what we can show, since our deployment had to manufacture the community it then remembered.

\subsubsection{An interactive system its community programs}

Participants discussed what they would want to try with the executable memory. W6 proposed a scavenger hunt for language learning, with hints written in the target language and learners in the same area grouped so that finding the objects is the exercise. W2 and W4 wanted collections that maintain themselves, channels that appear when a topic does and expire twenty minutes later, and notifications routed between named people on named subjects. W8 wanted the system to gather what their friends had been doing and narrate it back each night as fiction rather than as a summary, and separately to hold them to a daily plan. W13 wanted it to organise activities among people it had worked out share an interest, and to draw somebody's routine and say what they are probably doing an hour from now.

What these describe is a change of role. The memory would no longer be a thing that just records and recalls; it would be an agent, or a program, that holds the community's knowledge in order to act on it. How executable a shared memory should be, and what role it should then occupy in a community, is the open question this raises, and it is not one the architecture settles.

\subsubsection{A memory with an inside and an outside}

For our current prototype, everything reaches everyone, and a member cannot say that a post is for some people and not others. Participants wanted that shape, and wanted it by closeness rather than by topic, tiers running from a household to a floor to the whole campus (W12). W8 proposed an unusual way to build it, writing the permission layers inside the memory as ordinary posts rather than bolting them on as external access control, which would make scope something the community programs like any other behaviour. What this opens is not privacy alone but subject matter the deployment never saw: a memory a household or a project team could use for things they would not put in front of forty-six people, and sub-communities like the coffee and outdoor groups the wiki already discovered (Figure~\ref{fig:clouds}C) given their own layer rather than being flattened into one shared page. The cost is real, since a memory improves with contributors and scoping withholds exactly the material that would connect them, which we return to in Section~\ref{sec:discussion}.

\section{Discussion}\label{sec:discussion}

\subsection{Implementation Limitations}

\emph{\textbf{Recall is a strategy choice, and choosing badly loses connections silently.}} The system does not hold the whole memory in context when it composes a reply; it plans, retrieves and reconstructs, so material that exists can be missed, and nothing in a reply says which strategy ran or what another would have found. Participants noticed: several follow-ups ask outright what the search strategy was, or why an earlier sighting had been passed over. Prompting the planner to enumerate aspects more thoroughly would help, as would giving it a compressed view of the whole memory to plan against.

\emph{\textbf{It is not really multimodal.}} Images, video and audio enter the memory as generated descriptions and transcriptions rather than as media, so what the system knows about a photograph is whatever its caption says, and identity rests on wording, location and authorship rather than on seeing. Video is handled by describing a few frames, which makes anything recorded while moving poor quality. The consequences are concrete: one participant (W5) wanted to identify a person by their voice, which the architecture does not support at all; the same subject photographed from a different angle or in a different place may not be recognized as the same thing; and two objects that differ only slightly in appearance become indistinguishable once both are prose. Visual embeddings are present in the system, but they only retrieve candidate posts, and every judgment made about those candidates is made over descriptions. Closing this would mean carrying the media into the reasoning stages rather than only into retrieval, so that the reflection step can look rather than read.

\emph{\textbf{Knowledge is biased toward the heaviest contributors.}} A participant raised this during the study rather than us finding it, posting a question about whether a community's most prolific poster becomes its influencer and skews what the memory knows toward one person's view. The data suggests this concern is well founded: five of the 46 participants produced 51\% of the posts, with a single user contributing 21\% , while the median participant posted three times and 17 participants posted once. The memory has no representation of consensus, of how widely a belief is held, or  whose experience is missing, so a claim resting on one prolific contributor and a claim resting on twenty are indistinguishable in a reply. As W13 put it, “This is how social media works”: participation is uneven, and the resulting record can over-represent those who contribute most. Other participants noted that they often do not think to post at all (W6, W8), and argued for forms of capture that do not depend on remembering to contribute.

\subsection{Study limitations}

designed to observe what emerged rather than to measure system performance. A larger and longer study would be needed to test how knowledge accumulates over time and whether SEAM’s responses shape what participants later choose to post; here we can only describe that feedback loop as plausible. Subjective feedback was also limited: ratings were optional and sparse, and the follow-up workshop involved fourteen self-selected participants.

We likewise did not evaluate the memory quality itself: whether its conclusions were correct, whether its connections were useful, or whether it helped participants find information more effectively than a wiki, search interface, or colleague. Those questions require controlled comparisons beyond the scope of this deployment.

\subsection{Safety and consent}

The capabilities SEAM introduces also create corresponding risks.

\emph{\textbf{Reading the background is a powerful property but an exposing one.}} Recovering a detail from the background of a photograph was what some participants singled out as most striking, and also what they named as least comfortable (W1, W3, W4, W11). It may share sensitive information in a way the users were not aware of.

\emph{\textbf{An executable memory acts on people who did not consent to it.}} Instructions were written about third parties, one running on a participant's posts for eight days, and her only route out was to write an instruction of her own. A blanket request to disable all standing rules revoked only its author's.

\emph{\textbf{The memory may reveal more than any individual contribution.}} By combining posts, the knowledge layer can expose where people live, what they own, and where they are likely to be. Participants asked for greater control over who could see their posts and the inferences drawn from them. One design direction is therefore to expose abstractions while withholding the particulars that support them, for example describing what is known about a place without revealing the coordinates from which it was inferred.

\subsection{Deployment platform}

``Imagine this being done with the virtual goggles and all that. You just look at something and you literally see flashbacks, but from other people'' (W8). First-person capture removes the contribution cost that limits what the memory can hold, and an ambient or head-worn display lets recall arrive as something closer to a flashback than to a reply. The two requests came from opposite directions, easier capture asked for by several participants and the display proposed separately, which is some evidence that the pairing is the next platform for this rather than a gadget attached to it. Multimodal capture and multimodal recall then become one requirement. The clearest sketch of it in our data is a deposit left at a place: ``there is a bench you sit on, and it was an important moment. You say something, and then you sit on that bench again and you hear it back. Or you leave it for somebody else to hear'' (W8). That is a place-scoped executable memory with a different display, which this architecture already states and cannot yet deliver.

Such a surface sharpens the exposure and authority problems rather than resolving them, since recall arriving unasked in a visual field is harder to decline than a reply in a thread. That is what has to be built alongside the capability, not a reason to stop short of it.

\subsection{A shared mind is for acting together}

What participants described was closer to inheritance rather than retrieval: ``It is as if you have never seen this cat, but you look at it and you immediately know what it is called'' (W8). The deployment produced a weak form of that already, when somebody learns an animal's name and history from the posts of strangers. What matters is what follows the knowing. Everything the system itself can do stops at the reply and at the page it writes, so the consequential acting happens afterwards, in the world, by the person who now knows something nobody told them. The deployment contains one instance of that full loop, an instruction left at a place that reached a later participant who then acted on it physically, which makes the members of a memory its effectors rather than its audience. That a system in this position can shape behaviour is not speculative: altering what members saw in their feeds changed what they themselves went on to post, at scale and without their noticing, and the demonstration drew an editorial expression of concern about consent when it was published~\cite{kramer2014experimental}. A memory that acts through its members does openly what that experiment did covertly, which is what makes the capability credible and the responsibility sharper. We did not instrument any of this and treat it as a direction rather than a result. The fragment-assembly use described above is the same mechanism at a larger scale, a team each of whom sees part of a situation and is kept in step by a memory that relates their fragments. That is the version worth building toward: not everyone holding the same thoughts, but a group whose distributed knowledge produces coordinated action without first being gathered by any one person.

\bibliographystyle{ACM-Reference-Format}
\bibliography{references}

\end{document}